\pdfoutput=1
\documentclass[conference]{IEEEtran}
\IEEEoverridecommandlockouts
\usepackage{amsmath}
\usepackage{cite}
\usepackage{amssymb}
\usepackage{amsfonts}
\usepackage{algorithm}
\usepackage{dsfont}
\usepackage{graphicx}
\usepackage{epsfig}
\usepackage{subfigure}
\usepackage{psfrag}
\usepackage{xcolor}

\newtheorem{theorem}{Theorem}

\newcommand{\mv}[1]{\mbox{\boldmath{$ #1 $}}}

\title{{Intelligent Reflecting Surface Assisted Massive MIMO Communications}\thanks{The work was supported in part by the Key Area R\&D Program of Guangdong Province with grant No. 2018B030338001, by the National Key R\&D Program of China with grant No. 2018YFB1800800, by Natural Science Foundation of China with grant NSFC-61629101, and by Guangdong Zhujiang Project No. 2017ZT07X152.}}
\author{\IEEEauthorblockN{(\emph{Invited Paper})\\Zhaorui Wang${}^\dagger$, Liang Liu${}^\dagger$, and Shuguang Cui${}^\ddagger{}$}
	\IEEEauthorblockA{${}^\dagger$ EIE Department, The Hong Kong Polytechnic University. Email: \{zhaorui.wang,liang-eie.liu\}@polyu.edu.hk\\
		${}^\ddagger$ FNii and SRIBD, The Chinese University of Hong Kong, Shenzhen. Email: shuguangcui@cuhk.edu.cn}}
\begin{document}
\maketitle \thispagestyle{empty} \vspace{-0.3in}

\begin{abstract}
In a practical massive MIMO (multiple-input multiple-output) system, the number of antennas at a base station (BS) is constrained by the space and cost factors, which limits the throughput gain promised by theoretical analysis. This paper thus studies the feasibility of adopting the intelligent reflecting surface (IRS) to further improve the beamforming gain of the uplink communications in a massive MIMO system. Under such a novel system, the central question lies in whether the IRS is able to enhance the network throughput as expected, if the channel estimation overhead is taken into account. In this paper, we first show that the favorable propagation property for the conventional massive MIMO system without IRS, i.e., the channels of arbitrary two users are orthogonal, no longer holds for the IRS-assisted massive MIMO system, due to its special channel property that each IRS element reflects the signals from all the users to the BS via the same channel. As a result, the maximal-ratio combining (MRC) receive beamforming strategy leads to strong inter-user interference and thus even lower user rates than those of the massive MIMO system without IRS. To tackle this challenge, we propose a novel strategy for zero-forcing (ZF) beamforming design at the BS and reflection coefficients design at the IRS to efficiently null the inter-user interference. Under our proposed strategy, it is rigorously shown that even if the channel estimation overhead is considered, the IRS-assisted massive MIMO system can always achieve higher throughput compared to its counterpart without IRS, despite the fact that the favorable propagation property no longer holds.
\end{abstract}


\section{Introduction}\label{sec:Introduction}
Thanks to the properties of favorable propagation, i.e., user channels are orthogonal, and channel hardening, i.e., the strength of user channels does not fade over time, the massive MIMO (multiple-input multiple-output) technology is envisioned to be the key component in the fifth-generation (5G) cellular networks for improving the user throughput \cite{bjornson2016massive,ngo2013energy,marzetta2016fundamentals}. Theoretically speaking, the capacity of a massive MIMO system grows monotonically with the number of antennas at the base station (BS) due to the beamforming gain. However, in practice, the number of antennas at the BS is limited by the array dimensions allowed by the site owner, the weight, and the wind load. As a result, it remains an open problem in how to further reap the beamforming gain for improving the network throughput given the fact that it is practically difficult to deploy more than a few hundred of antennas per BS.

In this paper, we study the feasibility of adopting the intelligent reflecting surface (IRS) \cite{Liaskos08,Renzo19,Basar19} to further improve the throughput of the massive MIMO system. When the BS is equipped with a small number of antennas, the joint optimization of the BS beamforming vectors and IRS reflecting coefficients was studied in\cite{Wu18,Zhang19,Xu20,Shuowen20}, where the effectiveness of the IRS in enhancing the user signal-to-interference-plus-noise ratios (SINRs) was verified. However, in the IRS-assisted massive MIMO system, complicated optimization is not allowed considering the complexity issue in large systems. Moreover, network throughput depends on both the user SINRs and channel estimation overhead. However, the effect of the increased channel estimation overhead arising from the new channels related to the IRS \cite{wang2019channel_cof,wang2019channel} on the throughout is not taken into consideration in the above works. Therefore, it is necessary to revisit the role of IRS in massive MIMO systems if complicated optimization is infeasible and channel estimation overhead is considered.

This paper considers the uplink communications in a massive MIMO system, where an IRS is deployed to assist the single-antenna users to send their individual messages to the BS equipped with a large number of antennas.  First, we show analytically that the favorable propagation property no longer holds in the considered IRS-assisted massive MIMO system, because of the fact that each IRS element reflects the signals from all the users to the BS via the same channel. Thus, the maximal-ratio combining (MRC) beamforming can no longer cancel the inter-user interference and leads to lower user SINRs than those in the case without IRS. Next, to tackle this issue, we propose a novel design of the zero-forcing (ZF) beamforming vectors at the BS and reflection coefficients at the IRS. Different from \cite{Wu18,Zhang19,Xu20,Shuowen20} in which the IRS reflection coefficients are optimized based on the instantaneous channels, our scheme simply sets all these coefficients as one. In this case, the channel estimation overhead is exactly the same as the case without IRS, since the BS merely needs to estimate the users' effective channels to design the ZF solution. Moreover, we show rigorously that even if each IRS reflection coefficient is just set as one, the optimal ZF beamforming strategy can always yield higher user SINRs compared to the massive MIMO system without IRS. Since the channel estimation overhead is the same, it is concluded that the IRS-assisted massive MIMO system can always achieve higher throughput than its counterpart without the IRS.

\begin{figure}[t]
	\centering
	\includegraphics[width=8cm]{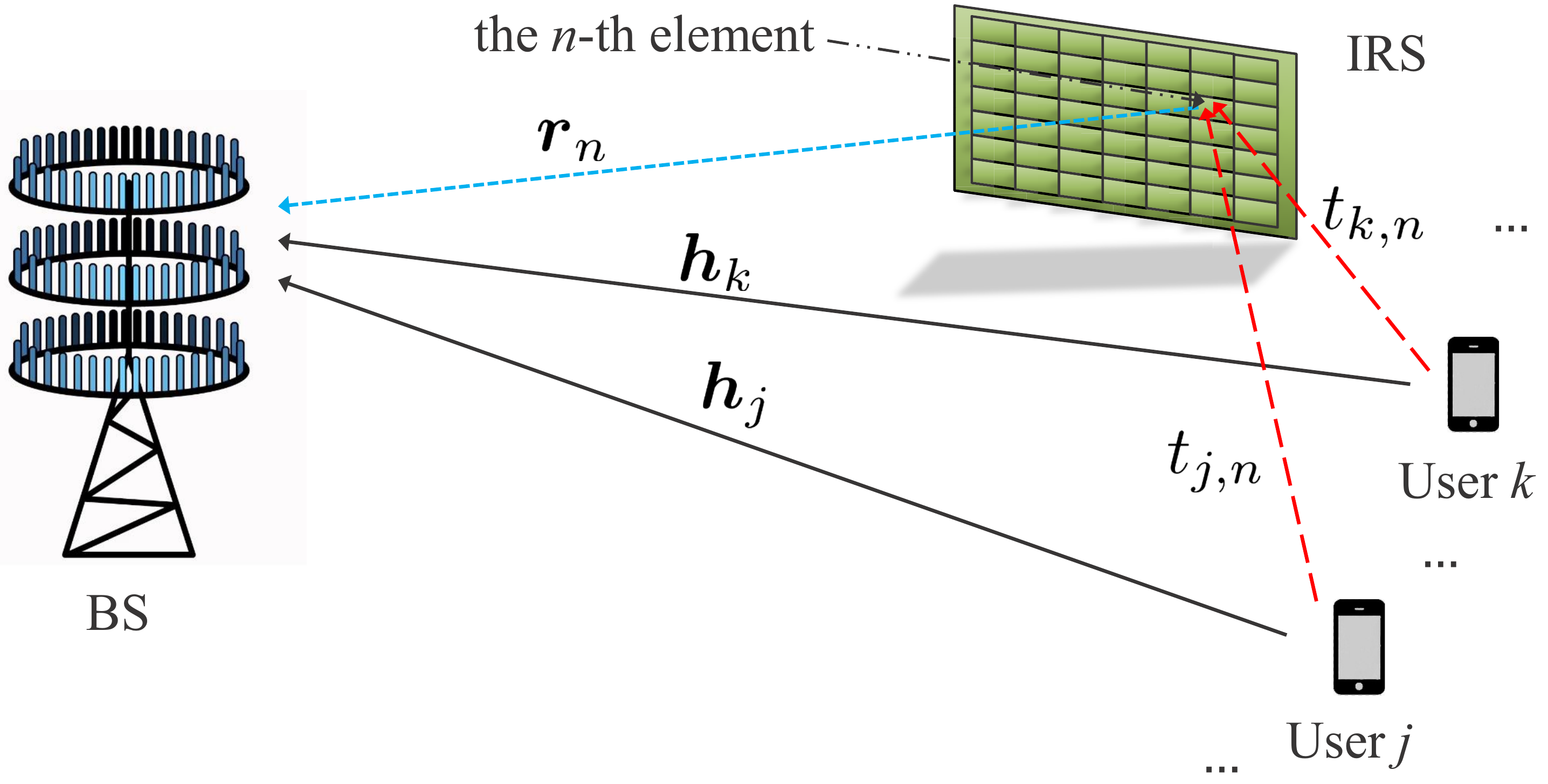}\vspace{-10pt}
	\caption{An IRS-assisted multiuser massive MIMO communication system.}\label{Fig1}\vspace{-10pt}
\end{figure}\vspace{-5pt}
\section{System Model}\label{sec:SYS}
We study a  massive MIMO system in which $K$ single-antenna users communicate to a BS with $M$ antennas in the uplink. An IRS  with $N\gg K$ passive reflecting elements is deployed to enhance the communication performance, as shown in Fig. \ref{Fig1}. Define $\phi_{n}$ with $|\phi_n|\leq 1$ as the reflection coefficient of the $n$th IRS element, $n=1,\cdots,N$. Through the IRS controller, the IRS elements are able to adjust $\mv{\phi}=[\phi_1,\cdots,\phi_N]^T$ to re-scatter the signals from the users.

We assume quasi-static block-fading channels, in which all channels remain approximately constant in each fading block with length $T$ symbols. Let $\mv{h}_k=(\mv{C}_k^{{\rm B}})^{\frac{1}{2}}\mv{h}_k^{{\rm IID}}\in\mathbb{C}^{M\times1}$, $k=1,\cdots,K$, denote the direct channel from the $k$th user to the BS, where $\mv{h}_k^{{\rm IID}}\sim \mathcal{CN}(\mv{0},\beta_k^{{\rm BU}}\mv{I})$, $\beta_k^{{\rm BU}}$ denotes the path loss of $\mv{h}_k$, and $\mv{C}_k^{{\rm B}}\succ \mv{0}$ with all diagonal elements being one denotes the BS receive correlation matrix for user $k$. Further, define $\mv{r}_n\in \mathbb{C}^{M\times 1}$ as the channel from the $n$th IRS element to the BS, $\forall n$, and $\mv{R}=[\mv{r}_1,\cdots,\mv{r}_N]$ as the overall channel from the IRS to the BS. We assume $\mv{R}=(\mv{C}^{{\rm B}})^{\frac{1}{2}}\mv{R}^{{\rm IID}}(\mv{C}^{{\rm I}})^\frac{1}{2}$, where $\mv{R}^{{\rm IID}}\sim \mathcal{CN}(\mv{0},M\beta^{{\rm BI}}\mv{I})$, $\beta^{{\rm BI}}$ denotes the path loss of $\mv{r}_n$'s, $\forall n$, and $\mv{C}^{{\rm B}}\succ\mv{0}$ and $\mv{C}^{{\rm I}}\succ\mv{0}$ with all diagonal elements being one denote the BS receive correlation matrix and IRS transmit correlation matrix for $\mv{R}$, respectively. Last, define $t_{k,n}\in \mathbb{C}$ as the channel from user $k$ to the $n$th IRS element, $\forall k,n$, and $\mv{t}_k=[t_{k,1},\cdots,t_{k,N}]^T$ as the overall channel from user $k$ to the IRS, $\forall k$. We assume that $\mv{t}_k=(\mv{C}_k^{{\rm I}})^{\frac{1}{2}}\mv{t}_k^{{\rm IID}}$, $\forall k$, where $\mv{t}_k^{{\rm IID}}\sim \mathcal{CN}(\mv{0},\beta_k^{{\rm IU}}\mv{I})$, $\beta_k^{{\rm IU}}$ denotes the path loss of $t_{k,n}$'s, $\forall n$, and $\mv{C}_k^{{\rm I}}\succ\mv{0}$ with all diagonal elements being one denotes the IRS receive correlation matrix for $\mv{t}_k$. In this paper, it is assumed that the BS uses $\tau<T$ symbols to estimate the useful channels. More information about the channel estimation time will be discussed in Sections III and IV.

Under the above model, the received signal at the BS is
\begin{align}
\mv{y}&=\sum_{k=1}^{K}\mv{h}_k\sqrt{p}s_k+\sum_{k=1}^{K}\sum_{n=1}^{N}\phi_{n}t_{k,n}\mv{r}_n\sqrt{p}s_k+\mv{z} \nonumber \\
&=\sum_{k=1}^{K}\left(\mv{h}_k+\sum_{n=1}^{N}\phi_{n}\mv{g}_{k,n}\right)\sqrt{p}s_k+\mv{z},
\label{eq:Sys-1.5}
\end{align}
where $s_k\sim \mathcal{CN}(0,1)$ denotes the transmit message of user $k$;  $\mv{z}\sim\mathcal{CN}\left(\mv{0},\sigma^2\mv{I}\right)$ denotes additive white Gaussian noise (AWGN) at the BS; $p$ denotes the identical transmit power of the users; and\vspace{-6pt}
\begin{align}
\mv{g}_{k,n}=t_{k,n}\mv{r}_n, ~~~ \forall n,k, \label{eq:Sys-2} \vspace{-4pt}
\end{align}
denotes the effective channel from the $k$th user to the BS through the $n$th IRS element. Note that for each IRS element $n$, the same $\mv{r}_n$ appears in all of $\mv{g}_{k,n}$'s, $k=1,\dots,K$.

The BS applies a linear beamforming vector $\mv{w}_k\in \mathbb{C}^{M\times 1}$ to decode $s_k$, $k=1,\dots,K$, i.e.,
\begin{align}
\mv{\hat{y}}_k=\sum_{j=1}^K\mv{w}_k^H\left(\mv{h}_j+\sum_{n=1}^{N}\phi_{n}\mv{g}_{j,n}\right)\sqrt{p}s_j\hspace{-2pt}+\hspace{-2pt}\mv{w}_k^H\mv{z}.
\end{align}
Then, the SINR for decoding $s_k$ is
\begin{align}
\gamma_k=\frac{p\left|\mv{w}_k^H\left(\mv{h}_k+\sum_{n=1}^{N}\phi_{n}\mv{g}_{k,n}\right)\right|^2}{p\sum\limits_{j\ne k}\left|\mv{w}_k^H\left(\mv{h}_j+\sum_{n=1}^{N}\phi_{n}\mv{g}_{j,n}\right)\right|^2\hspace{-2pt}+\sigma^2\mv{w}_k^H\mv{w}_k}. \label{eq:Sys-A1.7}
\end{align}
Moreover, the achievable rate of user $k$ is
\begin{align}
R_k=\frac{T-\tau}{T}\log_2(1+\gamma_k), ~~~ k=1,\cdot,K,
\label{eq:Sys-1.6}
\end{align}
where $\frac{T-\tau}{T}$ denotes the fraction of time for data transmission.
\vspace{-7pt}
\section{Fundamental Limitation of the MRC Strategy}\label{sec:limitation}
In a traditional massive MIMO system without the existence of an IRS, i.e., $\phi_n=0, ~\forall n$, the so-called favorable propagation property holds, i.e.,
\begin{align}
\lim_{M\to\infty}\frac{\left(\mv{h}_k\right)^H\mv{h}_j}{M}=0, ~~\forall j\ne k. \label{eq:fp2}
\end{align}
As a result, a simple MRC receiver, i.e., $\mv{w}_k=\mv{h}_k$, is optimal to maximize the SINR of user $k$, $\forall k$. According to \cite{ngo2013energy}, when $M\rightarrow \infty$ and the transmit power is set as $p=\frac{E}{M}$, where $E$ is fixed, the SINR of user $k$ with MRC converges to
\begin{align}
\bar{\gamma}_k^{\rm{(MRC,I)}}=\frac{E}{\sigma^2}\beta_k^{\text{BU}}, ~k=1,\dots,K. \label{eq: Sys-2}
\end{align}
Moreover, the MRC receivers require the knowledge of $\mv{h}_k$'s, which can be obtained via $\tau=K$ time slots in the channel training stage\cite{Hassibi03}. Thus, under the MRC receivers, the achievable rate of user $k$ when $M$ goes to infinity is
\begin{align}
\bar{R}_k^{\rm{(MRC,I)}}=\frac{T-K}{T}\log_2(1+\frac{E}{\sigma^2}\beta_k^{\text{BU}}), ~k=1,\dots,K. \label{eq:traditional-rate}
\end{align}

Motivated by the superiority of the MRC receivers in the conventional massive MIMO system, in the rest of this section, we study their performance in our considered IRS-assisted massive MIMO system. For convenience, define
\begin{align}\label{eqn:channel}
\mv{\hat{h}}_k=\mv{h}_k+\sum_{n=1}^{N}\phi_{n}\mv{g}_{k,n}, ~k=1,\dots,K,
\end{align}
as the effective channel between user $k$ and the BS. Then, the MRC beamforming vectors in the IRS-assisted system are $\mv{w}_k=\mv{\hat{h}}_k$,  $k=1,\dots,K$. With the above MRC receivers,  the SINRs given in \eqref{eq:Sys-A1.7} become
\begin{align} \gamma_k^{\rm{(MRC,II)}}=\frac{p\left|\mv{\hat{h}}_k^H\mv{\hat{h}}_k\right|^2}{\sum\limits_{j\ne k}p\left|\mv{\hat{h}}_k^H\mv{\hat{h}}_j\right|^2\hspace{-2pt}+\hspace{-2pt}\sigma^2\mv{\hat{h}}_k^H\mv{\hat{h}}_k},~k=1,\dots,K. \label{eq:III-3}
\end{align}
Under the considered channel model introduced in Section \ref{sec:SYS}, it can be shown that as $M\rightarrow \infty$, we have
\begin{align}
&\lim_{M\to\infty}\frac{\mv{\hat{h}}_k^H\mv{\hat{h}}_k}{M} =\beta_k^{\text{BU}}+ \beta^{\text{BI}}\mv{\eta}_k^H(\mv{\phi})\mv{C}^{{\rm I}}\mv{\eta}_k(\mv{\phi}), ~~~ \forall k, \label{eq:III-3.5} \\
&\lim_{M\to\infty}\frac{\mv{\hat{h}}_k^H\mv{\hat{h}}_j}{M} =\beta_k^{\text{BI}}\mv{\eta}_k^H(\mv{\phi})\mv{C}^{{\rm I}}\mv{\eta}_j(\mv{\phi})\neq 0, ~~~ \forall k\neq j, \label{eq:Sys-5.33}
\end{align}where $\mv{\eta}_k(\mv{\phi})=[\phi_1t_{k,1},\cdots,\phi_Nt_{k,N}]^T$, $\forall k$.

Different from (\ref{eq:fp2}), a key observation from \eqref{eq:Sys-5.33} is that in our considered IRS-assisted massive MIMO system, the favorable propagation property no longer holds. This is because as shown in \eqref{eq:Sys-2}, each IRS element reflects all the users' signals to the BS with the same channel, leading to channel correlation among different users. Based on \eqref{eq:III-3}, \eqref{eq:III-3.5}, and \eqref{eq:Sys-5.33}, we have the following theorem.
\begin{theorem}\label{theorem1}
	Assume that the transmit power of each user is $p=\frac{E}{M}$, where $E$ is fixed. Then, when $M$ goes to infinity, the SINR of user $k$, $k=1,\cdots,K$, achieved by any IRS reflection coefficients $\mv{\phi}$ and MRC receivers $\mv{w}_k=\mv{\hat{h}}_k$'s converges to
	\begin{align}
	&\bar{\gamma}_k^{({\rm MRC,II})}(\mv{\phi})=\lim\limits_{M\rightarrow \infty}\gamma_k^{\rm{(MRC,II)}}(\mv{\phi})=\nonumber\\
	&\frac{\left(\beta_k^{\text{BU}}+ \beta^{\text{BI}}\mv{\eta}_k^H(\mv{\phi})\mv{C}^{{\rm I}}\mv{\eta}_k(\mv{\phi})\right)^2E}{\sum\limits_{j\ne k}\left|\beta_k^{\text{BI}}\mv{\eta}_k^H(\mv{\phi})\mv{C}^{{\rm I}}\mv{\eta}_j(\mv{\phi})\right|^2\hspace{-2pt}E\hspace{-2pt}+\hspace{-2pt}\sigma^2\left(\beta_k^{\text{BU}}\hspace{-2pt}+ \hspace{-2pt} \beta^{\text{BI}}\mv{\eta}_k^H(\mv{\phi})\mv{C}^{{\rm I}}\mv{\eta}_k(\mv{\phi})\right)}. \label{eq:II-2}
	\end{align}
\end{theorem}

According to Theorem \ref{theorem1}, we need to optimize $\mv{\phi}$ to maximize the user SINRs, which requires the knowledge of  $\mv{g}_{k,n}$'s. Moreover, the design of the MRC receivers requires the knowledge of $\mv{h}_k$'s. All these channels can be estimated by $\tau=N+2K-1$ time slots using the method proposed in \cite{wang2019channel_cof} in a massive MIMO system. As a result, when $M$ goes to infinity, the rate of user $k$, $k=1,\cdots,K$, under the MRC receiver is
\begin{align}
\hspace{-2pt}\bar{R}_k^{\rm{(MRC,II)}}\hspace{-2pt}=\hspace{-2pt}\frac{T\hspace{-2pt}-\hspace{-2pt}N\hspace{-2pt}-\hspace{-2pt}2K\hspace{-2pt}+\hspace{-2pt}1}{T}\log_2\left(1+\bar{\gamma}_k^{\rm{(MRC,II)}}(\mv{\phi}^\ast)\right), \label{eq:IRS-MRC-rate}
\end{align}
where $\mv{\phi}^\ast$ denotes the optimal IRS reflection coefficients that can maximize the user SINRs in \eqref{eq:II-2}.

According to (\ref{eq:traditional-rate}) and (\ref{eq:IRS-MRC-rate}), it is observed that with IRS, the channel estimation overhead is enhanced. Moreover, the inter-user interference generally decreases the user SINRs, e.g., when $E/\sigma^2$ goes to infinity, $\bar{\gamma}_k^{\rm{(MRC,I)}}$ goes to infinity as well, but each $\bar{\gamma}_k^{({\rm MRC,II})}(\mv{\phi}^\ast)$ converges to a finite value. As a result, with IRS, the MRC receivers can even lead to lower user rates than the massive MIMO system without IRS.
\section{A Novel ZF-based Design}
In this section,  we first provide a simple solution of ZF beamforming design at the BS and reflecting coefficients design at the IRS to combat the fundament limits of the MRC-based solution shown in last section. Then, we rigorously show that under the propose solution, the IRS-assisted massive MIMO communication system can always achieve higher throughput than that achieved by its counterpart without IRS, as shown in (\ref{eq:traditional-rate}).

\subsection{Main Result}
In the proposed solution, we simply set the reflection coefficients of all the IRS elements as one, i.e.,
\begin{align}\label{eqn:unit phi}
\phi_n=1, ~~~ \forall n.
\end{align}Then, the effective channel between user $k$ and the BS shown in (\ref{eqn:channel}) reduces to
\begin{align}\label{eqn:new effective channel}
\mv{\hat{h}}_k=\mv{h}_k+\sum_{n=1}^{N}\mv{g}_{k,n}, ~k=1,\dots,K.
\end{align}

With the above effective user channels, we aim to design the optimal ZF beamforming to maximize each user's SINR. Specifically, the ZF beamforming vectors need to satisfy the following constraints: $\hat{\mv{h}}_j^H\mv{w}_k=0$, $\forall k\neq j$. Define $\hat{\mv{H}}_{-k}=[\hat{\mv{h}}_1,\cdots,\hat{\mv{h}}_{k-1},\hat{\mv{h}}_{k+1},\cdots,\hat{\mv{h}}_K]^H$, which constitutes all the effective channels except for $\hat{\mv{h}}_k$, $\forall k$. Then, the above ZF constraints can be expressed as $\hat{\mv{H}}_{-k}\mv{w}_k=\mv{0}$, $\forall k$.

Define the singular value decomposition (SVD) of $\hat{\mv{H}}_k$ as $\hat{\mv{H}}_k=\mv{X}_k\mv{\Lambda}_k[\bar{\mv{Y}}_k,\tilde{\mv{Y}}_k]^H$, $\forall k$, where $\mv{X}_k\in \mathbb{C}^{(K-1)\times (K-1)}$, $\mv{\Lambda}_k\in \mathbb{C}^{(K-1)\times M}$, $\bar{\mv{Y}}_k\in \mathbb{C}^{M\times (K-1)}$, and $\tilde{\mv{Y}}_k\in \mathbb{C}^{M\times (M-K+1)}$ with $\tilde{\mv{Y}}_k^H\tilde{\mv{Y}}_k=\mv{I}$. It can be shown that any ZF beamforming vector that satisfies $\hat{\mv{H}}_{-k}\mv{w}_k=\mv{0}$ can be expressed as $\mv{w}_k=\tilde{\mv{Y}}_k\tilde{\mv{w}}_k$, where $\tilde{\mv{w}}_k\in \mathbb{C}^{(M-K+1)\times 1}$. Given the above general form of the ZF beamforming vectors, the SINR of user $k$ given in (\ref{eq:Sys-A1.7}) reduces to
\begin{align}
\gamma_k^{({\rm ZF,I})}=\frac{p\tilde{\mv{w}}_k^H\tilde{\mv{Y}}_k^H\hat{\mv{h}}_k\hat{\mv{h}}_k^H\tilde{\mv{Y}}_k\tilde{\mv{w}}_k}{\sigma^2\tilde{\mv{w}}_k^H\tilde{\mv{w}}_k}, ~ \forall k.
\end{align}The optimal solution to maximize the above SINRs is $\tilde{\mv{w}}_k=\tilde{\mv{Y}}_k^H\hat{\mv{h}}_k$, $\forall k$. As a result, the optimal ZF beamforming vectors that can maximize the users' SINRs are
\begin{align}\label{eqn:optimal ZF}
\mv{w}_k^\ast=\tilde{\mv{Y}}_k\tilde{\mv{Y}}_k^H\hat{\mv{h}}_k, ~~~ \forall k.
\end{align}With the above ZF beamforming vectors and IRS reflection coefficients given in (\ref{eqn:unit phi}), the SINR of user $k$ is
\begin{align}\label{eqn:optimal ZF SINR}
\gamma_k^{({\rm ZF,I})}=\frac{p \|\tilde{\mv{Y}}_k^H\tilde{\mv{h}}_k\|^2}{\sigma^2}, ~~~ \forall k.
\end{align}

Note that in the above design, the IRS reflection coefficients are independent with the channels. Moreover, to design the optimal ZF beamforming vectors (\ref{eqn:optimal ZF}), the BS merely needs to know the user effective channels $\hat{\mv{h}}_k$'s, which can be estimated using $\tau=K$ time slots \cite{Hassibi03}. As a result, if the channel estimation time is taken into consideration, under the IRS reflection coefficients given in (\ref{eqn:unit phi}) and the optimal ZF beamforming vectors given in (\ref{eqn:optimal ZF}), the achievable rate of user $k$ in our considered IRS-assisted massive MIMO system is
\begin{align}\label{eqn:optimal rate ZF}
\hspace{-8pt}\bar{R}_k^{({\rm ZF,I})}=\frac{T-K}{T}\log_2\left(1+\bar{\gamma}_k^{({\rm ZF,I})}\right), ~ \forall k,
\end{align}where $\bar{\gamma}_k^{({\rm ZF,I})}=\lim_{M\rightarrow \infty}\gamma_k^{({\rm ZF,I})}$. In general, it is hard to get a closed-form expression of each $\bar{R}_k^{({\rm ZF,I})}$. Nevertheless, the following theorem states one key property of $\bar{R}_k^{({\rm ZF,I})}$'s.
\begin{theorem}\label{theorem3}
Under the IRS reflection coefficients given in (\ref{eqn:unit phi}) and the optimal ZF beamforming vectors given in (\ref{eqn:optimal ZF}), the achievable rates of all the users in our considered IRS-assisted massive MIMO system are always larger than those achieved in the massive MIMO system without IRS, i.e.,
\begin{align}\label{eqn:better throughput}
\bar{R}_k^{({\rm ZF,I})}>\bar{R}_k^{({\rm MRC,I})}, ~~~ \forall k,
\end{align}where $\bar{R}_k^{({\rm MRC,I})}$'s are given in (\ref{eq:traditional-rate}).
\end{theorem}

\subsection{Proof of Theorem \ref{theorem3}}
To prove Theorem \ref{theorem3}, in this subsection, we provide a suboptimal ZF beamforming solution given (\ref{eqn:unit phi}), which can always achieve higher user SINRs than those achieved in the massive MIMO system without IRS shown in (\ref{eq: Sys-2}). The above statement is sufficient to prove Theorem \ref{theorem3}, since the optimal ZF beamforming solution given in (\ref{eqn:optimal ZF}) can achieve higher SINRs than any suboptimal ZF solution. In the following, we show how to find such a suboptimal ZF beamforming solution.

In our proposed suboptimal solution, the received beamforming vectors are given by
\begin{align}
\mv{w}_k=\mv{h}_k+\sum_{n=1}^{N}\theta_{k,n}\mv{g}_{k,n},~ k=1,\dots,K,\label{eq: III-11}
\end{align}
where $\theta_{k,n}$'s are designed to null the interference, i.e.,
\begin{align}
\mv{w}_k^H\mv{\hat{h}}_j=\left(\mv{h}_k+\sum_{n=1}^{N}\theta_{k,n}\mv{g}_{k,n}\right)^H\mv{\hat{h}}_j=0,~\forall k\ne j. \label{eq:Sys-A1}
\end{align}
Note that \eqref{eq:Sys-A1} can be rewritten as
\begin{align}
\mv{A}_k\mv{\theta}_k=\mv{b}_k,~k=1,\dots,K, \label{eq:Sys-A4}
\end{align}
where $\mv{\theta}_k=\left[\theta_{k,1},\dots,\theta_{k,N}\right]^T$,
\begin{align}
&\mv{A}_k=
\left[\begin{array}{ccc}
\mv{\hat{h}}_1^H\mv{g}_{k,1}, & \cdots, & \mv{\hat{h}}_1^H\mv{g}_{k,N} \\ \vdots & \ddots & \vdots \\
\mv{\hat{h}}_{k-1}^H\mv{g}_{k,1}, & \cdots, & \mv{\hat{h}}_{k-1}^H\mv{g}_{k,N} \\
\mv{\hat{h}}_{k+1}^H\mv{g}_{k,1}, & \cdots, & \mv{\hat{h}}_{k+1}^H\mv{g}_{k,N} \\ \vdots & \ddots & \vdots \\
\mv{\hat{h}}_{K}^H\mv{g}_{k,1}, & \cdots, & \mv{\hat{h}}_{K}^H\mv{g}_{k,N} \\ \end{array} \right],\label{eq:Sys-A3}
\end{align}
and
\begin{align}
\mv{b}_k=-\left[\mv{\hat{h}}_1^H\mv{h}_k,\dots,\mv{\hat{h}}_{k-1}^H\mv{h}_k,\mv{\hat{h}}_{k+1}^H\mv{h}_k,\dots,\mv{\hat{h}}_{K}^H\mv{h}_k\right]^T. \label{eq:Sys-b}
\end{align}
Next, we consider the linear equations given in \eqref{eq:Sys-A4}. Define $\lambda_k={\rm{rank}}(\mv{A}_k)\leq K-1, \forall k$. Since we have $N>K$ unknown variables in $\mv{\theta}_k$, \eqref{eq:Sys-A4} describes an underdetermined system. As a result, there exist multiple solutions of $\mv{\theta}_k$'s to \eqref{eq:Sys-A4}. In this paper, we construct $\mv{\theta}_k$'s in the following way. Define the SVD of $\mv{A}_k/M$ as
\begin{align}
\frac{\mv{A}_k}{M}=\mv{U}_k\mv{\Sigma}_k \mv{V}_k^H, ~k=1,\dots,K. \label{eq:Sys-A4.5}
\end{align}
In \eqref{eq:Sys-A4.5}, $\mv{U}_k\in\mathbb{C}^{(K-1)\times(K-1)}$ and $\mv{V}_k^H=\left[\mv{v}_{k,1},\dots,\mv{v}_{k,N}\right]^H\in\mathbb{C}^{N\times N}$ are unitary matrices, and $\mv{\Sigma}_k\in\mathbb{C}^{(K-1)\times N}$ is expressed as
\begin{align}
\mv{\Sigma}_k=\left[\begin{array}{ccccccc}
\mv{\Sigma}_k^{(1)} &\mv{0}_{\lambda_k,N-\lambda_k} \\ \mv{0}_{K-1-\lambda_k,\lambda_k} &\mv{0}_{K-1-\lambda_k,N-\lambda_k}
\end{array}
\right], \label{eq:sigma}
\end{align}
where $\mv{0}_{i,j}$ is the all-zero matrix with dimension $i\times j$, $\mv{\Sigma}_k^{(1)}=$ ${\rm diag}([\delta_{k,1},\cdots,\delta_{k,{\lambda_{k}}}]^T)$ with ${\rm diag}(\mv{x})$ denoting a diagonal matrix whose diagonal entries are given by $\mv{x}$, and $\delta_{k,i}>0$'s, $i=1,\dots,\lambda_k$, are the positive singular values of $\mv{A}_k$.

Since $\mv{U}_k$ is a unitary matrix, \eqref{eq:Sys-A4} is equivalent to
\begin{align}
\mv{\Sigma}_k \mv{V}_k^H\mv{\theta}_k=\mv{U}_k^H\frac{\mv{b}_k}{M}, ~k=1,\dots,K. \label{eq:Sys-A4.6}
\end{align}
For convenience, define
\begin{align}
&\mv{\hat{\theta}}_k=\left[\hat{\theta}_{k,1},\dots,\hat{\theta}_{k,N}\right]^T=\mv{V}_k^H\mv{\theta}_k,\\
&\mv{\hat{b}}_k=\left[\hat{b}_{k,1},\dots,\hat{b}_{k,K-1}\right]^T=\mv{U}_k^H\frac{\mv{b}_k}{M}, ~~~ \forall k.
\end{align}
In addition, we define $\mv{\hat{\theta}}_k^{(1)}=\left[\hat{\theta}_{k,1},\dots,\hat{\theta}_{k,\lambda_k}\right]^T\in\mathbb{C}^{\lambda_k\times 1}$, $\mv{\hat{\theta}}_k^{(2)}=\left[\hat{\theta}_{k,\lambda_k+1},\dots,\hat{\theta}_{k,N}\right]^T\in\mathbb{C}^{(N-\lambda_{k})\times 1}$, and $\mv{\hat{b}}_k^{(1)}=\left[\hat{b}_{k,1},\dots,\hat{b}_{k,\lambda_{k}}\right]^T$. In this case, \eqref{eq:Sys-A4.6} requires
\begin{align}
\mv{\Sigma}_k^{(1)}\mv{\hat{\theta}}_k^{(1)}=\mv{\hat{b}}_k^{(1)},~k=1,\dots,K. \label{eq:Sys-A4.7}
\end{align}
We then have
\begin{align}
\mv{\hat{\theta}}_k^{(1)}=\left(\mv{\Sigma}_k^{(1)}\right)^{-1}\mv{\hat{b}}_k^{(1)}. \label{eq:Sys-A4.8}
\end{align}
Further, \eqref{eq:Sys-A4.6} holds given any $\mv{\hat{\theta}}_k^{(2)}$'s due to the structure of $\mv{\Sigma}_k$'s shown in \eqref{eq:sigma}. In this paper, we construct $\mv{\hat{\theta}}_k^{(2)}$'s in the following way. Define 
\begin{align}\label{eqn:Dk}
\mv{D}_k=(\mv{C}^{\rm I})^{\frac{1}{2}}{\rm diag}(\mv{t}_k)\mv{V}_k^{(2)}, ~~~ \forall k,
\end{align}where $\mv{V}_k^{(2)}=\left[\mv{v}_{k,\lambda_k+1},\dots,\mv{v}_{k,N}\right]$ is a sub-matrix of $\mv{V}_k$ consisting of the last $N-\lambda_k$ columns of $\mv{V}_k$. Moreover, define the QR decomposition of $\mv{D}_k$ as $\mv{D}_k=\mv{Q}_k\mv{R}_k$, $\forall k$, where $\mv{Q}_k\in \mathbb{C}^{N\times (N-\lambda_k+1)}$ satisfies $\mv{Q}_k^H\mv{Q}_k=\mv{I}$, and $\mv{R}_k\in \mathbb{C}^{(N-\lambda_k+1)\times (N-\lambda_k+1)}$ is an upper triangular matrix. Then, $\mv{\hat{\theta}}_k^{(2)}$'s are given as
\begin{align}
\mv{\hat{\theta}}_k^{(2)}=\mv{R}_k^{-1}\mv{Q}_k^H(\mv{C}^{\rm{I}})^{\frac{1}{2}}\mv{t}_k, ~~~ \forall k. \label{eq:Sys-B0}
\end{align}

At last, given $\hat{\mv{\theta}}_k=\left[\left(\hat{\mv{\theta}}_k^{(1)}\right)^T,\left(\hat{\mv{\theta}}_k^{(2)}\right)^T\right]^T$'s, under the ZF beamforming design in \eqref{eq: III-11}, we can set
\begin{align}
\mv{\theta}_k=\mv{V}_k\mv{\hat{\theta}}_k, ~~~ k=1,\dots,K.\label{eq:III-theta}
\end{align}

\begin{theorem}\label{theorem2}
	Assume that the transmit power of each user is  $p=\frac{E}{M}$. Then, when $M$ goes to infinity, by setting the IRS reflection coefficients as $\phi_n=1,~ \forall n$, the SINR of user $k$ achieved by the ZF beamforming vectors in \eqref{eq: III-11} and \eqref{eq:III-theta} converges to
	\begin{align}
	&\bar{\gamma}_k^{({\rm ZF},{\rm II})}=\lim\limits_{M\rightarrow \infty}\gamma_k^{({\rm ZF},{\rm II})}\nonumber\\
	=&\bar{\gamma}_k^{\rm{(MRC,I)}}+\frac{E}{\sigma^2}\beta^{\rm BI}\mv{t}_k^H(\mv{C}^{\rm{I}})^{\frac{1}{2}}\mv{Q}_k\mv{Q}_k^H(\mv{C}^{\rm{I}})^{\frac{1}{2}}\mv{t}_k, ~\forall k, \label{eq:Sys-B5.6}
	\end{align}where $\bar{\gamma}_k^{\rm{(MRC,I)}}$ is the SINR of user $k$ achieved in the massive MIMO system without IRS, as given in (\ref{eq: Sys-2}).
\end{theorem}

\begin{IEEEproof}
Please refer to Appendix \ref{appendix2}.
\end{IEEEproof}


Theorem \ref{theorem2} indicates that if the IRS reflection coefficients are set as $\phi_n=1$, $\forall n$, and the ZF beamforming vectors are set as \eqref{eq: III-11} and \eqref{eq:III-theta}, we have
\begin{align}
\bar{\gamma}_k^{({\rm ZF},{\rm II})}> \bar{\gamma}_k^{\rm{(MRC,I)}}, ~~~ \forall k.
\end{align}Moreover, the ZF beamforming solution given in \eqref{eq: III-11} and \eqref{eq:III-theta} is sub-optimal. As a result, given the optimal ZF beamforming solution (\ref{eqn:optimal ZF}), it follows that
\begin{align}
\bar{\gamma}_k^{({\rm ZF,I})}\geq \bar{\gamma}_k^{({\rm ZF},{\rm II})}> \bar{\gamma}_k^{\rm{(MRC,I)}}, ~~~ \forall k.
\end{align}Theorem \ref{theorem3} is thus proved.

\section{Numerical Examples}\label{sec:Numerical Examples}
In this section, we provide a numerical example to compare the network throughput achieved by the following three schemes: IRS-assisted massive MIMO under the optimal ZF beamforming (\ref{eqn:optimal ZF}), IRS-assited massive MIMO under the MRC beamforming, and massive MIMO without IRS under MRC beamforming.  We assume that the BS is equipped with $M=512$ antennas, the number of users is $K=8$, and the number of IRS elements ranges from $10\le N\le 210$. The transmit power is $p=\frac{E}{M}$, where $E$ is set to be 13 dBm in all setups.  The channel bandwidth is assumed to be $100$ MHz, and the power spectrum density of the AWGN is $-169$ dBm/Hz. The fading block length is $T=1000$. In addition, we consider the exponential correlation matrix model \cite{wang2019channel} to characterize $\mv{C}_k^{{\rm B}}$'s, $\mv{C}^{{\rm B}}$, $\mv{C}^{{\rm I}}$, and $\mv{C}_k^{{\rm I}}$'s.

\begin{figure}[t]
	\centering
	\includegraphics[width=9cm]{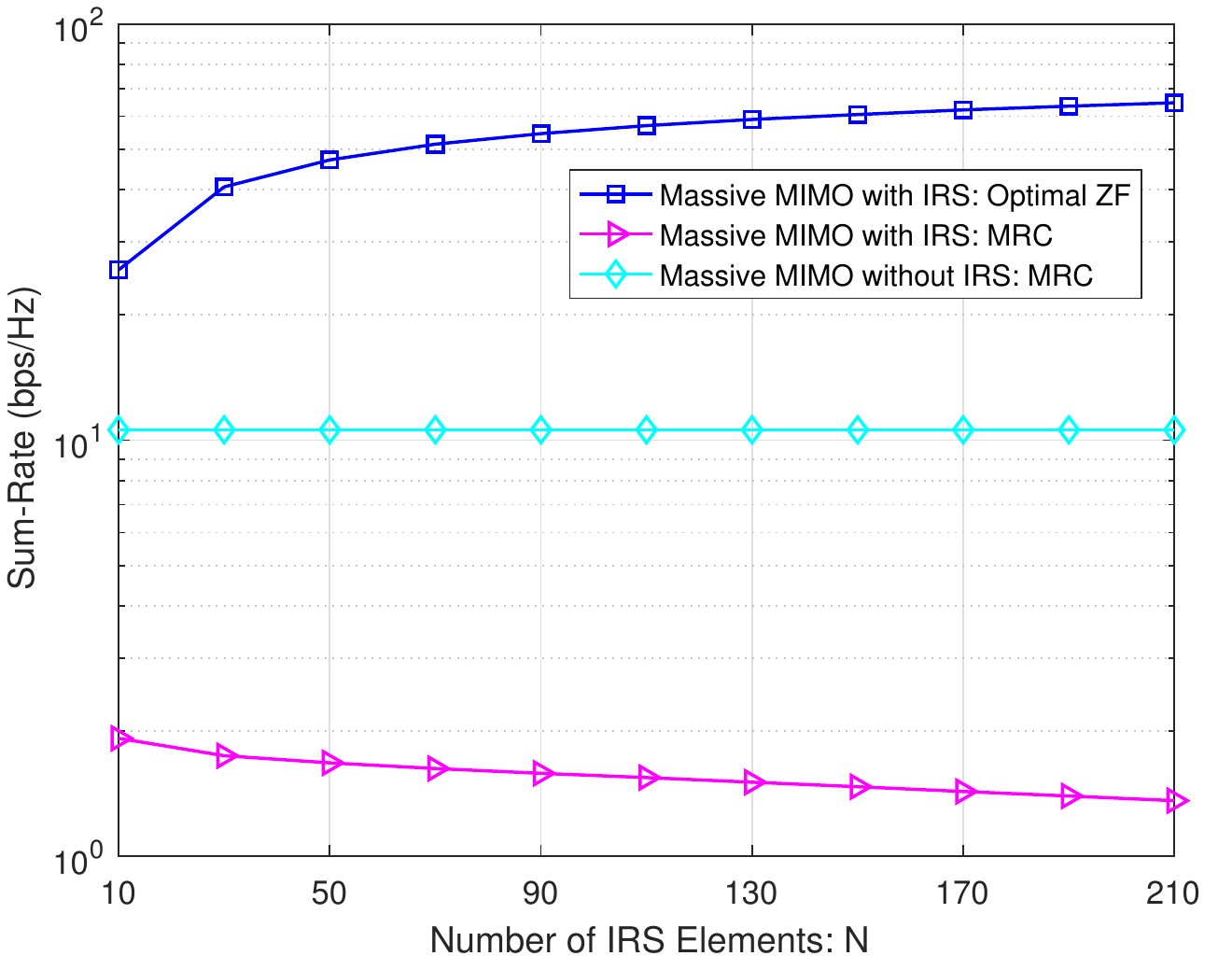}\vspace{-10pt}
	\caption{Sum-rate comparison between the three schemes.}\label{Fig2}\vspace{-10pt}
\end{figure}

Fig. \ref{Fig2} shows the sum-rate performance of the 8 users versus different numbers of IRS elements under the three schemes. First, it is observed that in the IRS-assisted massive MIMO system, the sum-rate performance of the MRC receivers is even worse than that in the massive MIMO system without IRS, as indicated in Section \ref{sec:limitation}. Next, it is observed that under the IRS reflection strategy given in (\ref{eqn:unit phi}) and the optimal ZF beamforming strategy given in (\ref{eqn:optimal ZF}), the sum-rate achieved in the IRS-assisted massive MIMO system is much larger than that achieved in the massive MIMO system without IRS, as predicted in Theorem \ref{theorem3}. Moreover, the sum-rate gain becomes larger as the number of IRS elements increases.

\section{Conclusion}
This paper studied the feasibility of applying IRS to further improve the user rates in the massive MIMO systems when the channel estimation overhead is taken into consideration. We first showed that in IRS-assisted massive MIMO systems, the MRC beamforming cannot null the inter-user interference as in the conventional massive MIMO systems without IRS. To address this issue, we proposed a novel design of ZF beamforming vectors at the BS and reflection coefficients at the IRS to efficiently null the interference. Under this solution, we showed rigorously that the IRS-assisted massive MIMO system can always use the same channel estimation overhead to achieve higher user SINRs compared to its counterpart without IRS. As a result, IRS is very effective in improving the user rates in the massive MIMO systems.

\begin{appendix}
\subsection{Proof of Theorem \ref{theorem3}}\label{appendix2}
In \eqref{eq:Sys-A1.7}, by setting $p=\frac{E}{M}$, $\mv{w}_k$'s as \eqref{eq: III-11} and  \eqref{eq:III-theta}, we have
\begin{align}
&\bar{\gamma}_k^{({\rm ZF},{\rm II})}=\lim_{M\to\infty}\gamma_k^{({\rm ZF},{\rm II})}=\lim_{M\to\infty}\frac{E\left|\mv{w}_k^H\mv{\hat{h}}_k\big{/}M\right|^2}{\sigma^2\mv{w}_k^H\mv{w}_k/M}\nonumber\\
&=\!\!\lim_{M\to\infty} \!\!\frac{\left|\left(\mv{h}_k\!+\!\sum_{n=1}^{N}\!\!\theta_{k,n}\mv{g}_{k,n}\right)^H\!\!\!\left(\mv{h}_k\hspace{-2pt}\!+\!\hspace{-2pt}\sum\limits_{n=1}^{N}\mv{g}_{k,n}\right)\!\!\Big{/}\!M\!\right|^2E}{\hspace{-2pt}\!\!\sigma^2\left(\mv{h}_k\!+\!\sum_{n=1}^{N}\!\theta_{k,n}\mv{g}_{k,n}\right)^H\!\!\!\left(\mv{h}_k\!+\!\sum_{n=1}^{N}\!\theta_{k,n}\mv{g}_{k,n}\right)\!\!\Big{/}\!M} \nonumber \\
&=\frac{\left|\beta_k^{\text{BU}}+\beta^{\text{BI}}\lim\limits_{M\to\infty}\mv{\theta}_k^H({\rm{diag}}(\mv{t}_k))^H\mv{C}^{\rm{I}}\mv{t}_k\right|^2E}{\hspace{-2pt}\sigma^2\left(\beta_k^{\text{BU}}+\beta^{\text{BI}}\lim\limits_{M\to\infty}\mv{\theta}_k^H({\rm{diag}}(\mv{t}_k))^H\mv{C}^{\rm{I}}{\rm{diag}}(\mv{t}_k)\mv{\theta}_k\right)},~\forall k, \label{eq:Sys-B2-4}
\end{align}
where \eqref{eq:Sys-B2-4} follows from the fact that $\gamma_k^{({\rm ZF},{\rm II})}$'s are continuous functions. It can be shown from \eqref{eq:Sys-b} that as $M$ goes to infinity, we have
\begin{align}
\lim\limits_{M\rightarrow \infty} \frac{\mv{b}_k}{M}=\mv{0}, ~~~ \forall k.
\end{align}As a result, it follows from \eqref{eq:Sys-A4} that
\begin{align}
\lim_{M\to\infty}\frac{\mv{A}_k}{M}\mv{\theta}_k=\lim_{M\to\infty}\frac{\mv{b}_k}{M}=\mv{0},~k=1,\dots,K, \label{eq:APD3}
\end{align}
where
\begin{align}
\lim_{M\to\infty}\frac{\mv{A}_k}{M}=
\beta^{\text{BI}}
\left[\begin{array}{ccc}
\mv{a}_{1,k}^T\\
\vdots\\
\mv{a}_{k-1,k}^T\\
\mv{a}_{k+1,k}^T\\
\vdots\\
\mv{a}_{K,k}^T\\ \end{array} \right]\ne \mv{0}, ~\forall k, \label{eq:APD4}
\end{align}
with $\mv{a}_{j,k}^T=\mv{t}_j^H\mv{C}^{\rm I}{\rm diag}(\mv{t}_k)$, $\forall j\neq k$. Similar to \eqref{eq:Sys-A4.5}-\eqref{eq:Sys-A4.7}, it can be shown that as $M$ goes to infinity, we have
\begin{align}
\lim_{M\to\infty}\mv{\Sigma}_k^{(1)}\mv{\hat{\theta}}_k^{(1)}=\lim_{M\to\infty}\mv{\hat{b}}_k^{(1)}=\mv{0},~\forall k.\label{eq:APDsigma}
\end{align}
Moreover, since $\lim_{M\to\infty}\mv{A}_k/M\neq \mv{0}$, $\lambda_k=\rm{rank}(\lim_{M\to\infty}\mv{A}_k/M)\neq 0$. As a result, $\lim_{M\to\infty}\mv{\Sigma}_k^{(1)}$ is a full rank matrix. \eqref{eq:APDsigma} then indicates
\begin{align}
\lim\limits_{M\to\infty}\mv{\hat{\theta}}_k^{(1)}=\mv{0}. \label{eq:Sys-A4.881}
\end{align}

With (\ref{eq:Sys-A4.881}), (\ref{eqn:Dk}), and (\ref{eq:Sys-B0}), it follows that
\begin{align}
& \lim\limits_{M\rightarrow \infty} (\mv{C}^{{\rm I}})^{\frac{1}{2}}{\rm diag}(\mv{t}_k)\mv{\theta}_k \nonumber \\
= & \lim\limits_{M\rightarrow \infty} (\mv{C}^{{\rm I}})^{\frac{1}{2}}{\rm diag}(\mv{t}_k)\mv{V}\hat{\mv{\theta}}_k \\
= & (\mv{C}^{{\rm I}})^{\frac{1}{2}}{\rm diag}(\mv{t}_k)\mv{V}^{(2)}\hat{\mv{\theta}}_k^{(2)} \\
= & \mv{Q}_k\mv{R}_k\mv{R}_k^{-1}\mv{Q}_k^H(\mv{C}^{{\rm I}})^{\frac{1}{2}}\mv{t}_k \\
= & \mv{Q}_k\mv{Q}_k^H(\mv{C}^{{\rm I}})^{\frac{1}{2}}\mv{t}_k, ~~~ \forall k.
\end{align}

As a result, the following condition holds
\begin{align}
& \lim\limits_{M\to\infty}\mv{\theta}_k^H({\rm{diag}}(\mv{t}_k))^H\!\mv{C}^{\rm{I}}\mv{t}_k \nonumber \\
= & \lim\limits_{M\to\infty}\mv{\theta}_k^H({\rm{diag}}(\mv{t}_k))^H\mv{C}^{\rm{I}}{\rm{diag}}(\mv{t}_k)\mv{\theta}_k \label{eqn:Appendix1}\\
= & \mv{t}_k^H(\mv{C}^{{\rm I}})^{\frac{1}{2}}\mv{Q}_k\mv{Q}_k^H(\mv{C}^{{\rm I}})^{\frac{1}{2}}\mv{t}_k, ~~~ \forall k. \label{eqn:Appendix2}
\end{align}

By taking (\ref{eqn:Appendix1}) and (\ref{eqn:Appendix2}) into (\ref{eq:Sys-B2-4}), Theorem \ref{theorem2} is thus proved. 

\end{appendix}

\bibliographystyle{IEEEtran}
\bibliography{CIC}

\end{document}